

\font\titlefont = cmr10 scaled\magstep 4
\font\sectionfont = cmr10

\font\teenyfont = cmr5

\magnification = 1200

\global\baselineskip = 1.2\baselineskip
\global\parskip = 4pt plus 0.3pt
\global\nulldelimiterspace = 0pt

\predisplaypenalty 1000


\def\endignore{}
\def\ignore #1\endignore{}

\newcount\dflag
\dflag = 0


\def\monthname{\ifcase\month
\or Jan \or Feb \or Mar \or Apr \or May \or June%
\or July \or Aug \or Sept \or Oct \or Nov \or Dec
\fi}




\def\endid{}
\def\id#1\endid{\number\day\ \monthname \number\year
\hfill #1}

\def\endtitle{}
\def\title#1\endtitle{\vskip.3in\titlefont
\global\baselineskip = 2\baselineskip
#1\vskip.4in
\baselineskip = 0.5\baselineskip\rm}

\def\lblfoot{This work was supported by the Director, Office of Energy
Research, Office of High Energy and Nuclear Physics, Division of High
Energy Physics of the U.S. Department of Energy under Contract
DE-AC03-76SF00098.}

\def\endauthors{}
\def\authors#1\endauthors{
#1\if\dflag = 0
\footnote{}{\noindent\lblfoot}\fi}

\def\endabstract{}
\def\abstract#1\endabstract{\vskip .3in%
\centerline{\sectionfont\bf Abstract}%
\vskip .1in%
\noindent#1%
\ifnum\dflag = 0
\footline = {\hfil}\pageno = 0
\vfill\eject
\pageno = 1\footline{\centerline{\sectionfont\folio}}
\fi\ifnum\dflag = 2
\footline = {\hfil}\pageno = 0
\vfill\eject
\fi}


\newcount\nsection
\newcount\nsubsection

\def\section#1{\global\advance\nsection by 1
\global\nsubsection = 0
\bigskip\noindent\sectionfont \bf \number\nsection.\ #1
\nobreak\medskip\rm\nobreak}

\def\subsection#1{\global\advance\nsubsection by 1
\bigskip\noindent\sectionfont \it \number\nsection.\number\nsubsection.\ #1%
\nobreak\medskip\rm\nobreak}

\def\appendix#1#2{\bigskip\noindent%
\sectionfont \bf Appendix #1.\ #2
\nobreak\medskip\rm\nobreak}


\newcount\nref
\global\nref = 1

\def\ref#1#2{\xdef #1{[\number\nref]}
\ifnum\nref = 1\global\xdef\therefs{\noindent[\number\nref] #2\ }
\else
\global\xdef\oldrefs{\therefs}
\global\xdef\therefs{\oldrefs\vskip.1in\noindent[\number\nref] #2\ }%
\fi%
\global\advance\nref by 1
}

\def\listrefs{\vfill\eject\section{References}\therefs}


\newcount\nfig
\global\nfig = 1

\def\fg#1\efig{\vskip .5in\noindent Fig.\ \number\nfig:\ #1%
\global\advance\nfig by 1}


\newcount\cflag
\newcount\nequation
\global\nequation = 1
\def\eqlabel{(1)}

\def\nexteqno{\ifnum\cflag = 0
\global\advance\nequation by 1
\fi
\global\cflag = 0
\xdef\eqlabel{(\number\nequation)}}

\def\lasteqno{\global\advance\nequation by -1
\xdef\eqlabel{(\number\nequation)}}

\def\label#1{\xdef #1{(\number\nequation)}
\ifnum\dflag = 1
{\escapechar = -1
\xdef\draftname{\teenyfont\string#1}}
\fi}

\def\clabel#1#2{\xdef\eqlabel{(\number\nequation #2)}
\global\cflag = 1
\xdef #1{\eqlabel}
\ifnum\dflag = 1
{\escapechar = -1
\xdef\draftname{\string#1}}
\fi}

\def\cclabel#1#2{\xdef\eqlabel{#2)}
\global\cflag = 1
\xdef #1{\eqlabel}
\ifnum\dflag = 1
{\escapechar = -1
\xdef\draftname{\string#1}}
\fi}


\def\eeq{}

\def\eqnn #1\eeq{$$ #1 $$}

\def\eq #1\eeq{\xdef\draftname{\ }
$$ #1
\eqno{\eqlabel \rlap{\ \draftname}} $$
\nexteqno}







\def\eqa #1\eeq{\xdef\draftname{\ }
$$ \eqalignno{ #1 } $$
\global\cflag = 0}





\def\jref#1#2#3#4{{\it #1} {\bf #2}, #3 (#4)}

\def\NPB#1#2#3{\jref{Nucl.\ Phys.}{B#1}{#2}{#3}}
\def\PA#1#2#3{\jref{Physica}{#1A}{#2}{#3}}
\def\PLB#1#2#3{\jref{Phys.\ Lett.}{#1B}{#2}{#3}}
\def\PR#1#2#3{\jref{Phys.\ Rep.}{#1}{#2}{#3}}
\def\PRD#1#2#3{\jref{Phys.\ Rev.}{D#1}{#2}{#3}}

\def\PRL#1#2#3{\jref{Phys.\ Rev.\ Lett.}{#1}{#2}{#3}}
\def\PRV#1#2#3{\jref{Phys.\ Rev.}{#1}{#2}{#3}}


\def\gotoo{\mathop{\longrightarrow}}


\def\frac#1#2{{{#1} \over {#2}}\,}  
\def\sfrac#1#2{{\textstyle\frac{#1}{#2}}}  

\def\ppartial#1#2{{{\partial #1} \over {\partial #2}}}  

\def\Dsl{\hbox{/\kern-.6000em\rm D}} 



\def\mybar#1{\kern 0.8pt\overline{\kern -0.8pt#1\kern -0.8pt}\kern 0.8pt}
\def\sla#1{\raise.15ex\hbox{$/$}\kern-.57em #1}
\def\Sla#1{\kern.15em\raise.15ex\hbox{$/$}\kern-.72em #1}

\def\roughly#1{\mathrel{\raise.3ex\hbox{$#1$\kern-.75em%
    \lower1ex\hbox{$\sim$}}}}

\def\scr#1{{\cal #1}}


\def\del{\delta}


\def\tr{\mathop{\rm tr}}



\def\avg#1{\langle #1 \rangle}




\def\GeV{{\rm \ GeV}}
\def\TeV{{\rm \ TeV}}

\id
LBL-32893, UCB-PTH-92/34
\endid

\title
\centerline{Technicolor Theories with Negative S}
\endtitle

\authors
\centerline{Markus A. Luty$^*$\ \ {\it and}\ \ Raman Sundrum$^{*\dagger}$}
\footnote{}{This work was supported in part by the Director, Office of Energy
Research, Office of High Energy and Nuclear Physics, Division of High Energy
Physics of the U.S. Department of Energy under contract DE-AC03-76SF00098 and
in part by the National Science Foundation under grant PHY90-21139.}
\vskip .1in
\centerline{\it $^*\,$Theoretical Physics Group}
\centerline{\it Lawrence Berkeley Laboratory}
\centerline{\it 1 Cyclotron Road}
\centerline{\it Berkeley, California 94720}\vskip .1in
\vskip .1in
\centerline{\it $^\dagger\,$Department of Physics}
\centerline{\it University of California}
\centerline{\it Berkeley, California 94720}
\endauthors

\abstract
We show that the pseudo Nambu--Goldstone boson contribution to the
Peskin--Takeuchi electroweak parameter $S$ can be negative in
a class of technicolor theories.
This negative contribution can be
large enough to cancel the positive techni-hadron contribution, showing that
electroweak precision tests alone
cannot be used to rule out technicolor as the mechanism of electroweak
symmetry breaking.
\vskip .1in\noindent
PACS numbers 12.15.Cc, 14.80.Gt
\endabstract


\def\GEW{$SU(2)_W \times U(1)_Y$}
\def\GTC{$SU(N)_{TC}$}
\def\GWW{$SU(2)_W \times SU(2)_{W'}$}


\ref\TC{See E. Farhi and L. Susskind \PR{74}{277}{1981} and references
therein.}

\ref\TCmodels{
M. Einhorn and D. Nash, Santa Barbara preprint NSF--ITP--91--91;
R. Sundrum, LBL preprint LBL-32107 (1992), to appear in
{\it Nucl.\ Phys.} {\bf B};
L. Randall, MIT preprint MIT-CTP\#2112 (1992);
H. Georgi, Harvard preprint \#HUTP-92/A037.}

\ref\TCrad{R. Renken and M. Peskin, \NPB{211}{93}{1983};
B. W. Lynn, M. Peskin and R. G. Stuart in
J. Ellis and R. D. Peccei {\it eds.}, {\it Physics at LEP} (1986);
M. Golden and L. Randall, \NPB{361}{3}{1991};
 A. Dobado, D. Espriu and M.
Herrero, \PLB{255}{405}{1990}; R. Johnson, B. L. Young and D.
W. McKay, \PRD{43}{R17}{1991}.}

\ref\HT{B. Holdom and J. Terning, \PLB{247}{88}{1990}.}

\ref\PT{M. E. Peskin and T. Takeuchi, \PRL{65}{964}{1990};
\PRD{44}{3641}{1992}.}

\ref\KL{D. C. Kennedy and B. W. Lynn, \NPB{322}{1}{1989}.}

\ref\effL{J.\ Schwinger, \PLB{24}{473}{1967};
S.\ Coleman, J.\ Wess, and B.\ Zumino, \PRV{117}{2239}{1969};
C.\ G.\ Callan, S.\ Coleman, J.\ Wess, and B.\ Zumino, \PRV{117}{2247}{1969};
S.\ Weinberg, \PA{96}{327}{1979}.}

\ref\negS{H. Georgi, \NPB{363}{301}{1991};
E. Gates and J. Terning, \PRL{67}{1840}{1991};
E. Ma and P. Roy, \PRL{68}{2879}{1992}.}

\ref\DR{M. J. Dugan and L. Randall, \PLB{264}{154}{1991}.}

\ref\Z{W. Marciano and J. Rosner, \PRL{65}{2963}{1990}; B. Holdom,
\PLB{259}{329}{1991}; P. Langacker and M. Luo, \PRD{45}{278}{1992}.}

\ref\CU{P. Sikivie, L. Susskind, M. Voloshin and V. Zakharov,
\NPB{173}{189}{1980}.}

\ref\Ncontrov{R.\ S.\ Chivukula, M. J. Dugan, and M. Golden, Boston University
preprint BUHEP--92--25 (Harvard preprint HUTP--92/A033).}

\ref\GL{J. Gasser and H. Leutwyler, \NPB{250}{1985}{465}.}

\ref\SS{M. Soldate and R. Sundrum, \NPB{340}{1}{1990}.}


\section{Introduction}

Technicolor \TC\ is a leading candidate for the mechanism of electroweak
symmetry breaking.
In technicolor  models, strong gauge forces give rise to a fermion condensate
which spontaneously breaks the electroweak symmetry similarly
to the spontaneous breaking of chiral symmetry in QCD.
Technicolor models solve the hierarchy problems associated with the standard
model in an elegant way, although to date they have failed to provide a
compelling explanation of the origin of fermion masses without
phenomenologically unacceptable flavor-changing neutral currents.
(For some recent attempts at realistic technicolor model building, see
refs.\ \TCmodels.)
While there is a widespread feeling that technicolor is unattractive, the
possibility that electroweak symmetry is broken by technicolor can only
be excluded by testing predictions of technicolor which are independent of
the details of any specific technicolor model.
Such tests may be forthcoming at the LHC and SSC, where techni-hadron
resonances are expected in the TeV mass range.

Recently, several groups have studied radiative corrections to electroweak
observables
in the context of technicolor theories \TCrad\HT\PT\ in an attempt to find
other
model-independent tests of technicolor.
These groups concluded that technicolor models generically give rise to large
deviations from standard model predictions.
In particular, refs.\ \HT\PT\ argued that the quantity $S$ (defined
below) is positive in technicolor models.
This is an important claim, since current data favors a negative value for
$S$ \PT.

However, we will show that positive $S$ is not a generic feature of
technicolor models by constructing simple technicolor theories in which
the contribution to $S$ from pseudo Nambu--Goldstone bosons (PNGB's) is large
and negative.
For reasonable estimates of the techni-hadron contribution to $S$, we
conclude that $S$ can be negative in these models.
These models are intended as existence proofs to show that the value of $S$
does not provide a
model-independent test of technicolor as the mechanism of electroweak symmetry
breaking.

The plan of this paper is as follows.
In section 2, we briefly discuss the $S$ parameter in technicolor theories.
In section 3, we give several examples of models in which the PNGB
contribution to $S$ can be negative.
Section 4 contains our conclusions.

\section{$S$ in Technicolor Theories}

Following ref.\ \PT, we define
\eq
S \equiv -16\pi \left. \ppartial{\Pi_{3Y}(q^2)}{q^2} \right|_{q^2 = 0},
\eeq
where $\Pi_{3Y}$ is the coefficient of $g_{\mu\nu}$ in the $W_3$--$Y$
vacuum polarization.
This is one of three parameters which completely characterize the ``oblique''
radiative corrections \KL\ due to physics at energies large compared to $M_Z$.

To discuss $S$ in technicolor models we will use an effective field theory
description.
For definiteness, we will focus on technicolor theories with technicolor group
\GTC\ with $K$ technifermions in the fundamental representation.
At a scale $\sim 1 \TeV$, technicolor becomes strong and breaks the
technifermion chiral symmetry in the pattern
\eq
SU(K)_L \times SU(K)_R \gotoo SU(K)_{L + R},
\eeq
giving rise to $K^2 - 1$ NGB's.
Three of these NGB's become the longitudinal components of the $W$ and $Z$
bosons, while the remainder are assumed to get masses from a combination of
electroweak gauge interactions and ``extended technicolor'' interactions (see
below).

The interactions of the NGB's can be described by a nonlinear effective
lagrangian \effL.
The NGB fields are described by a $K \times K$ traceless hermitian matrix
field $\Pi$
transforming under $SU(K)_L \times SU(K)_R$ as
\eq
\Sigma(x) = e^{i\Pi(x) / f} \mapsto L \Sigma(x) R^\dagger,
\eeq
where $f$ is the NGB decay constant.
The \GEW\ covariant derivative acting on $\Sigma$ is
\eq
\label\covd
D_\mu \Sigma = \partial_\mu \Sigma
+ ig_2 W_{\mu a} \left( T_{La} \Sigma - \Sigma T_{Ra} \right)
+ ig_1 B_\mu \left( \sfrac 12 Y_L \Sigma - \Sigma \sfrac 12 Y_R \right),
\eeq
where $T_{L, R}$ ($Y_{L, R}$) are the generators of $SU(2)_W$ ($U(1)_Y$) acting
on the left- and right-handed technifermions.
The effective lagrangian is
\eq
\label\theL
\scr L = \frac{f^2}{4} \tr\left(D^\mu \Sigma^\dagger D_\mu \Sigma \right)
+ \cdots.
\eeq
The effective lagrangian contains terms that give
rise to $S \ne 0$ at tree level, such as
\eq
\del\scr L = \frac{\sigma}{16\pi^2}
\tr\left( [D^\mu, D^\nu] \Sigma [D_\mu, D_\nu] \Sigma^\dagger \right).
\eeq
Terms such as this encode the contribution to $S$ from physics above the cutoff
of the effective lagrangian, which we will refer to as the ``hadronic''
contribution.

Examples of new physics which can give rise to negative contributions to $S$
have been given \negS\DR, but these do not seem to work well in technicolor.
Elementary scalars can be invoked, but this conflicts with the main motivation
for technicolor, namely solving the hierarchy problem without
supersymmetry.
Heavy fermions can give rise to negative $S$, but large electroweak symmetry
violating masses are required, which are typically difficult to obtain in
technicolor theories. Also the negative contributions to $S$ obtained in this
way are small unless many fermions are present, so it is difficult to cancel
the positive hadronic contribution.
Finally, extensions of the electroweak gauge group can give rise to large
radiative corrections \Z, but these are not of the ``oblique'' type, and
we do not consider them here.
We are therefore led to consider the contribution to $S$ from technicolor
itself, in particular PNGB's.
The couplings of the PNGB's are determined by the technifermion representations
and are therefore highly constrained.
Nonetheless, we find that simple models exist in which the PNGB
contributions to $S$ can be large and negative.

\section{Technicolor Models with $S < 0$}

We will consider technicolor models with the following features:

$\bullet$ {\it Custodial symmetry} {\rm \CU}:
We will assume that the theory possesses an approximate unbroken $SU(2)_C$
custodial symmetry which protects the relation $\rho = 1$.
The custodial symmetry is broken by $U(1)_Y$ interactions, and also by the
mechanism which gives rise to fermion mass splittings.
Radiative corrections from custodial symmetry violation will not be
discussed here.

We will implement custodial symmetry in a simple way by assuming that
$U(1)_Y$ is imbedded in a group $SU(2)_{W'} \times U(1)_X$ via
$\sfrac 12 Y = T_3' + X$,
where $T_3'$ ($X$) is a $SU(2)_{W'}$ ($U(1)_X$) generator.
These groups are imbedded into the chiral symmetries of the technifermions so
that the symmetry breaking pattern is
\eq
SU(2)_W \times SU(2)_{W'} \times U(1)_X \gotoo SU(2)_C \times U(1)_X,
\eeq
where $SU(2)_C = SU(2)_{W + W'}$ is the custodial symmetry.
The photon is massless, since $Q = T_3 + T'_3 + X$ is unbroken.

$\bullet$ {\it Extended technicolor:}
A realistic technicolor model must contain additional interactions in order to
generate fermion masses.
We will refer to these new interactions as ``extended technicolor'' (ETC),
although we do not assume that they are necessarily due to massive gauge boson
exchange.
We will assume that any new particles associated with the ETC sector are
heavier than $\sim 1 \TeV$.
These new interactions will in general give rise to a complicated mass
spectrum for the PNGB's.
It is this feature that we will exploit to construct models with $S < 0$.

$\bullet$ {\it Techni-hadron dynamics:}
We will be focussing on the PNGB contribution to $S$, but of course the
measured value of $S$ also contains the contribution from techni-hadrons.
The size of this contribution for the case of many technifermions is
currently somewhat controversial \Ncontrov.
However we have made simple estimates that indicate that there is no
inconsistency in using large-$N$ counting to estimate this contribution,
even when the number of technifermions is large.
We will therefore use large-$N$ counting consistent with the estimates of
refs.\ \HT\PT\ and comment on the uncertainties when we present our results.

The leading PNGB contribution to $S$ comes from a single loop of PNGB's and is
easily evaluated from the interactions contained in the term eq.\ \theL:
\eq
\label\sngb
S_{\rm PNGB} = \frac{1}{24\pi} \, \sum_{r,s} \gamma_{rs}\, I(m_r, m_s),
\eeq
where the sum is over the $SU(2)_C$ representations of the PNGB's;
$I$ is the kinematic factor
\eq
I(m_r, m_s) = 6 \int_0^1 dx\, x(1 - x)\,
\ln\frac{\mu^2}{x m_r^2 + (1 - x) m_s^2},
\eeq
where $m_r$ is the mass of the PNGB's in representation $r$ of $SU(2)_C$ and
$\mu$ is taken to be the mass of the techni-$\rho$ obtained
from scaling up QCD: $\mu \simeq (3.5 \TeV) / \sqrt{N}$.
(Our results do not depend sensitively on this choice.)
$\gamma_{rs}$ is a group-theory factor
\eq
\gamma_{rs} = \tr\left( P_r\, T_{3V}\, P_s\, \sfrac 12 Y_V \right),
\eeq
where $T_{3V} = T_{3L} + T_{3R}$, {\it etc\/}.
(See eq.\ \covd.)
The trace is over the space of NGB's and $P_r$ is
the projection operator for the $SU(2)_C$ representation $r$.

\subsection{Model 1}

Consider a model with technifermions transforming under \GWW\ as
\eq
\eqalign{
\psi_L & \sim (\sfrac 12,\ 0), \qquad
\psi_R \sim (0,\ \sfrac 12), \cr
\chi_L & \sim (j,\ 0), \qquad\,\,
\chi_R \sim (j,\ 0). \cr}
\eeq
Here we have used ``spin'' labels for the $SU(2)$ representations.
All fermions are singlets under $U(1)_X$.
(By invoking custodial symmetry, it is easy to see that $S$ is independent of
any $U(1)_X$ quantum numbers we might assign to the fermions.)
The strong technicolor dynamics gives rise to condensates
$\avg{\mybar\psi_L \psi_R}$, $\avg{\mybar\chi_L \chi_R} \ne 0$
which break \GWW\ in the desired pattern;
the $\chi$ condensate does not contribute to electroweak breaking.

The NGB's in this model are easily classified.
We write the NGB field as
\eq
\Pi = \pmatrix{w & \xi \cr \xi^\dagger & \eta \cr}
\sim \pmatrix{\mybar\psi i\gamma_5 \psi & \mybar\chi i\gamma_5 \psi \cr
\mybar\psi i\gamma_5 \chi & \mybar\chi i\gamma_5 \chi \cr},
\eeq
where $w^\dagger = w$, $\eta^\dagger = \eta$.
The 3 NGB's in the $w$ block become the longitudinal components of the $W$ and
$Z$, while the other components are physical PNGB's.
The  PNGB's in the $\eta$ block do not carry hypercharge, and
therefore do not contribute to $S$.
The  PNGB's in the $\xi$ block belong to the $SU(2)_C$
representations $j \pm \frac 12$
Finally, there is an ``axion'' whose generator is proportional to the identity
matrix in both the $w$ and $\eta$ blocks, which also does not contribute to
$S$.

The PNGB's in the $\xi$ block give rise to the group theory factors
\eq
\eqalign{
\gamma_{j-\frac 12, j-\frac 12} &= -\frac{j(4j + 3)(2j - 1)}{3(2j + 1)}, \cr
\gamma_{j+\frac 12, j+\frac 12} &= \frac{(4j + 1)(2j + 3)(j + 1)}
{3(2j + 1)}, \cr
\gamma_{j+\frac 12, j - \frac 12} = \gamma_{j - \frac 12, j + \frac 12}
&= -\frac{4j(j + 1)}{3(2j + 1)}. \cr}
\eeq
If the PNGB's with smaller custodial spin are light, they tend to give
negative $S$ due to the chiral logarithm in eq.\ \sngb.
A similar pattern was noted in ref.\ \DR\ for elementary scalars.
In table 1, we show the resulting contributions to $S$ for a particular
choice of the PNGB mass spectrum.

\vbox{
\vskip 20pt
\centerline{
\vbox{\offinterlineskip
\hrule
\halign{&\vrule#&\strut\quad\hfil#\quad\cr
height2pt&\omit&&\omit&&\omit&&\omit&&\omit&\cr
&\omit\hfil&&$j = 2\ \ \ $&&$j = 5/2\ \ \ $&&$j = 3\ \ \  $&\cr
height2pt&\omit&&\omit&&\omit&&\omit&&\omit&\cr
\noalign{\hrule}
height1pt&\omit&&\omit&&\omit&&\omit&&\omit&\cr
\noalign{\hrule}
height2pt&\omit&&\omit&&\omit&&\omit&&\omit&\cr
&$S_{\rm PNGB}$&&$-0.10 \pm 0.02$&&$-0.20 \pm 0.06$&&$-0.34 \pm 0.15$&\cr
height2pt&\omit&&\omit&&\omit&&\omit&&\omit&\cr}
\hrule}}
{\leftskip=30pt\rightskip=30pt\noindent
Table 1: PNGB contributions to $S$ in model 1.
We have taken $N = 2$ and assumed that the PNGB's with custodial spin
$j \pm \frac 12$ have a mass of $1 \TeV$ ($200 \GeV$).
Using the estimate of the techni-hadron contribution from
ref.\ \PT, we obtain $S_{\rm had} \simeq 0.17$ for this model.
The quoted errors are our estimates of the contributions to $S$ from higher
orders in chiral pertrubation theory (see text), and are meant to show that
chiral perturbation theory has not broken down for $S$.
The main source of uncertainty in the total $S$ is the rescaling of strong
interaction quantities from QCD in the estimate of $S_{\rm had}$.
\par}
\vskip 20pt}

The PNGB mass splittings arise from several sources.
Electroweak interactions will give rise to PNGB masses at order
$g^2$, summarized in the effective lagrangian by the term
\eq
\label\EWmass
\del \scr L_{EW} = a g_2^2 f^4 \tr\left(
T_{aL} \Sigma T_{aR} \Sigma^\dagger \right).
\eeq
($f = 246 \GeV$ in this model.)
This term is $SU(2)_C$ invariant and will give the PNGB's in the $\xi$ block
masses
\eq
m^2_\ell = a g_2^2 v^2 \left[
j(j + 1) - \sfrac 34 + \ell(\ell + 1) \right],
\qquad \ell = j \pm \sfrac 12.
\eeq
By scaling up from QCD, we find $a g_2^2 f^2 \simeq (250 \GeV)^2 / N$.
Note the the electroweak splittings do make the larger custodial
representations heavier, but this does not give rise to significant negative
values for $S$.
We neglect order $g^4$ terms which are expected to give rise to custodial
symmetry violating PNGB mass splittings of less than $1$ GeV.

However, ETC interactions can also contribute to PNGB splittings.
For example, we can add to the theory the four-fermion interactions
\eq
\del\scr L_\pm = \frac 1{M_\pm^2}\,
\left( \mybar\psi_L P^\pm\chi_R \right)
\left( \mybar\psi_L P^\pm\chi_R \right)^\dagger,
\eeq
where $P^\pm$ are the projection operators onto the $SU(2)_W$
representation $j \pm \frac 12$.
These operators give masses to the $\xi$ PNGB's
\eq
m_{j\pm\frac 12}^2 \sim \frac{b f^4}{M_\pm^2},
\eeq
where $b \simeq 10 N^2$ by scaling up QCD.
For $M_\pm \sim 1 \TeV$, this can give rise to the PNGB mass splittings of
table 1.

Subleading ETC and electroweak contributions to $S$ can be large.
The leading corrections come from modifications of the couplings of
gauge bosons to PNGB's, and give rise to logarithmically enhanced contributions
to $S$ through PNGB loops.
Both ETC and electroweak corrections can be estimated to be
\eq
\frac{\delta S}S \sim \frac{m_P^2}{\Lambda^2},
\eeq
where $m_P$ is the contribution to the PNGB mass and $\Lambda$ is the
scale which acts as the expansion parameter in the low-energy effective
lagrangian.
We have taken $\Lambda \simeq (4.2 \TeV) / \sqrt{K}$, which is obtained by
scaling up QCD using the fact that corrections due to the
$s$ quark mass are expected to be of order $30\%$ \GL.
The dependence on the number of technifermions $K$ is dictated by demanding
that higher order terms be comparable to the size of loop corrections
\SS.

For the larger values of $j$ in table 1, strong ETC operators are
required to keep the $j - \frac 12$ NGB's light despite the large
electroweak mass contribution.
$M_-$ must therefore be fine tuned.
For $j = 3$  this is a $10\%$  fine tuning.

In the models of this paper, $SU(2)_W$ is not asymptotically free.
The maximum value of $j$ we considered was determined by requiring that
the energy at which $SU(2)_W$ becomes strong (estimated by the one-loop Landau
pole) be greater than $\sim 10$ TeV.

\vfil\eject
\subsection{Model 2}

In this model, the technifermions transform as
\eq
\eqalign{
\psi_L & \sim (\sfrac 12,\ 0), \qquad
\psi_R \sim (0,\ \sfrac 12), \cr
\chi_L & \sim (j,\ \sfrac 12), \qquad\,\,
\chi_R \sim (j,\ \sfrac 12). \cr}
\eeq
There are now many custodial PNGB representations which in general can have
different masses.
As before, we find that PNGB's with small custodial spin tend to give
rise to negative contributions to $S$.

To illustrate this, we choose a simple form of the PNGB spectrum.
Note that the $\chi$ fermions appear in custodial representations
$j \pm \frac 12$.
We assume that the dominant effect of ETC and electroweak interactions on the
PNGB spectrum is described below the electroweak scale by independent mass
terms for the
$\chi$ fermions in the different custodial representations.
(This can be arranged using gauge- and $SU(2)_C$- invariant six- and
four-fermion operators.
For $j = 2$  this requires a $10\%$  fine tuning in order to keep the PNGB's
in smaller $SU(2)_C$ representations light.)
PNGB's containing the $j + \sfrac 12$ fermions can then be made heavier
than those containing $j - \frac 12$ fermions, giving rise to negative values
of $S$, as illustrated in table 2.

\vbox{
\vskip 20pt
\centerline{
\vbox{\offinterlineskip
\hrule
\halign{&\vrule#&\strut\quad\hfil#\quad\cr
height2pt&\omit&&\omit&&\omit&&\omit&&\omit&\cr
&\omit\hfil&&$j = 1$&&$j = 3/2$&&$j = 2$&\cr
height2pt&\omit&&\omit&&\omit&&\omit&&\omit&\cr
\noalign{\hrule}
height1pt&\omit&&\omit&&\omit&&\omit&&\omit&\cr
\noalign{\hrule}
height2pt&\omit&&\omit&&\omit&&\omit&&\omit&\cr
&$S_{\rm PNGB}^{(1)}$&&$-0.5 \pm 0.2$&&$-1.5 \pm 0.5$&&$-3.2 \pm 1.8$&\cr
&$S_{\rm PNGB}^{(2)}$&&$-0.19 \pm 0.02$&&$-0.78 \pm 0.06$&&$-1.8 \pm 0.2$&\cr
height2pt&\omit&&\omit&&\omit&&\omit&&\omit&\cr}
\hrule}}
{\leftskip=30pt\rightskip=30pt\noindent
Table 2: PNGB contributions to $S$ in model 2.
We have taken $N = 4$.
The values in the first (second) row are for the case where the PNGB's
containing two $j + \frac 12$ fermions have a mass of  $1 \TeV$ ($500 \GeV$)
and the PNGB's containing two $j - \frac 12$ fermions have a mass of
$200 \GeV$.
Using the estimate of the techni-hadron contribution from
ref.\ \PT, we obtain $S_{\rm had} \simeq 0.34$ for this model.
As in model 1, the quoted errors are our estimates of the contributions to $S$
from higher orders in chiral pertrubation theory, and does not include the
uncertainty in $S_{\rm had}$.
\par}
\vskip 20pt}

The value of the $\rho$ parameter may be a problem for this model, since
there are custodial symmetry violating contributions to PNGB masses of order
$20 \GeV$ in the lighter PNGB's from $\,U(1)_Y$ gauge boson exchange.
There will also be contributions from ETC-induced mass splittings.
Simple estimates indicate that without large cancellations, this will give rise
to contributions to $\delta \rho$ close to the current experimental limits,
but this is highly dependent on the details of the ETC sector.
The fact that $S$ can be large even for a custodial symmetry invariant PNGB
mass spectrum such as that chosen (for simplicity) in table 2 shows that
$S$ can be large independently of $\delta \rho$ in this model.

\section{Conclusions}

We have shown that $S$ can be negative in simple technicolor theories.
The new ingredients in these models are the use of technicolor sectors which
do not break electroweak symmetry and a particular choice of higher-dimension
(``ETC'') interactions.
ETC-induced mass splittings of pseudo Nambu--Goldstone bosons arising from
these sectors can give rise to large negative values of $S$, cancelling the
positive ``hadronic'' contribution.
These models serve as illustrative existence proofs that experimental
constraints on $S$ cannot be used to rule out technicolor
as the mechanism of electroweak symmetry breaking.
Even ``large'' technicolor theories with many technifermions are not
excluded, since the negative contributions from PNGB's can be substantial.
Therefore, direct searches for particles associated with technicolor (pseudo
Nambu--Goldstone bosons and technihadrons) are still required to test
technicolor.

\section{Acknowlegements}

We thank M. Suzuki for discussions on the topic of this paper;
we thank L. Dixon, and especially M. J. Dugan and M. Golden for helpful
criticisms of an earlier draft.
This work was supported in part by the Director, Office of Energy Research,
Office of High Energy and Nuclear Physics, Division of High Energy Physics of
the U.S. Department of Energy under contract DE-AC03-76SF00098 and in part by
the National Science Foundation under grant PHY90-21139.

\listrefs

\bye